\documentclass[aps,pra,notitlepage,%
superscriptaddress,onecolumn]{revtex4-1}

\usepackage{natbib}
\usepackage{graphicx}
\usepackage{amssymb}
\usepackage{amsmath}
\usepackage{bm}
\usepackage{bbm}
\usepackage{comment}
\usepackage{float}
\usepackage{dcolumn}
\newcolumntype{d}[1]{D{.}{.}{#1}}

\usepackage{fancyvrb}

\newcommand{\dd}{\mathrm{d}}

\newcommand{\ee}{\mathrm{e}}

\usepackage{bbm}

\begin{document}

\title{Short--Range Hard--Sphere Potential and Coulomb Interaction:\\
Deser--Trueman Formula for Rydberg States of Exotic Atomic Systems}

\author{Gregory S. Adkins}
\affiliation{Department of Physics and Astronomy, Franklin \& Marshall College,
Lancaster, Pennsylvania 17604, USA}

\author{Ulrich D. Jentschura}
\affiliation{Department of Physics and LAMOR, Missouri University of Science and
Technology, Rolla, Missouri 65409, USA}

\begin{abstract}
In exotic atomic systems with hadronic constituent 
particles, it is notoriously difficult to estimate
the strong-interaction correction to energy levels. 
It is well known that, due to the strength of the 
nuclear interaction, the problem cannot be treated
on the basis of Wigner--Brillouin perturbation 
theory. Recently, high-angular-momentum Rydberg states of exotic atomic systems
with hadronic constituents have been identified as promising candidates 
for the search of New Physics in the low-energy 
sector of the Standard Model. 
We thus derive a generalized Deser--Trueman formula for 
the induced energy shift for a general hydrogenic 
bound state with principal quantum number $n$
and orbital angular momentum quantum number~$\ell$, and find that the energy shift
is given by the formula
$\delta E = 2 \alpha_{n,\ell} \, \beta_\ell \, (a_h/a_0)^{2 \ell + 1} E_h/n^3$,
where $\alpha_{n,0} = 1$, $\alpha_{n,\ell} = \prod_{s = 1}^\ell 
(s^{-2} - n^{-2})$, $\beta_\ell = (2 \ell + 1)/[(2 \ell + 1)!!]^2$,
where $E_h$ is the Hartree energy, 
$a_h$ is the hadronic radius and $a_0$ is the 
generalized Bohr radius. The square of the double factorial,
$[(2\ell + 1)!!]^2$, in the denominator implies a 
drastic suppression of the effect for higher angular momenta.
\end{abstract}

\maketitle

%
%
\section{Introduction}

Recently, the non--$S$ Rydberg states of exotic hadronic
atomic systems have been proposed as candidates 
for the search for New Physics effects as 
additions to the low-energy sector~\cite{BaEtAl2025,AdJe2025prr}.
Hence, the focus in the current work 
is on the strong-interaction energy correction
for bound states of hadronic bound systems with nonvanishing angular momenta,
where the strong-interaction shift is so small
that it is otherwise neglected in many treatments of 
the exotic systems (see Sec.~5 of Ref.~\cite{GaLyRu2008}).
However, in the hunt for New Physics,
it is still necessary to estimate the 
strong-interaction shift because it can be an obstacle
to the detection of New Physics effects in the 
extreme electromagnetic fields encountered in the 
exotic hadronic bound systems~\cite{BaEtAl2025,AdJe2025prr}.

Examples of exotic bound systems 
(see also the comprehensive discussion in Appendix~\ref{appa})
studied in the literature include antiprotonic atoms~\cite{BaEtAl2025},
as well as $K$-mesic atoms (see Ref.~\cite{Tr1961}) 
such as those involving a negatively charged $K$ meson and a proton
($K^-$--p) or deuteron ($K^-$--d),
and protonium, the bound system of a proton 
and its antiparticle (see Refs.~\cite{CaRiWy1992,BaFrGa2001,KlBaMaRi2002}).
The DIRAC collaboration at CERN has focused
on the ground-state lifetime of pionium
(the $\pi^+\pi^-$ atom, see Refs.~\cite{DI2005,DI2015}),
and the bound pion-kaon system~\cite{DI2017}. 
The pionic hydrogen collaboration
at the Paul--Scherrer--Institute (PSI) has focused
on pionic hydrogen measurements (see Ref.~\cite{pionicH})
and pionic deuterium (Ref.~\cite{HaEtAl1998}).
The DEAR collaboration (DA$\Phi$NE Exotic Atom Research) 
at LNF/INFN has investigated kaonic hydrogen~\cite{BeEtAl2005},
including transitions from higher excited states of
kaonic hydrogen to the ground state (the so-called $K_\alpha$,
$K_\beta$ and $K_\gamma$ lines, see Ref.~\cite{BeEtAl2005}).
In general, the search for New Physics 
effects~\cite{SaEtAl2018} on the basis of Rydberg states
of two-body exotic hadronic atomic systems 
is very promising because the 
extreme field strengths experienced at a distance 
equal to the generalized Bohr radius
can exceed the Schwinger critical field strength,
(for muonic bound systems, see
Fig.~3 of Ref.~\cite{Je2015muonic}).
Thus, while, strictly speaking, ``Rydberg states''
can refer to any excited state of a bound Coulomb quantum 
system, we here focus on states with nonvanishing angular
momenta and, in particular, have in mind so-called ``circular
Rydberg states'' with the maximum angular momentum for 
given principal quantum number, and nearly circular states. Indeed,
the new approach outlined in 
Refs.~\cite{BaEtAl2025,AdJe2025prr}
focuses on non-$S$ states which provide us with 
an almost ideal scenario of extreme electric field
strength, combined with much reduced nuclear-size
and strong-interaction effects.
It is of crucial importance to 
estimate the effects due to the strong interaction
for the non-$S$ states. The strong
interaction becomes dominant at distances 
commensurate with $a_h = 1 \, {\rm fm}$.
A universally applicable,
approximate formula for the estimation 
of the strong-interaction 
effects~\cite{Tr1961,CaRiWy1992,BaFrGa2001,KlBaMaRi2002}
is thus highly desirable. This necessitates
a generalization of the Deser--Trueman 
formulas~\cite{DeGoBaTh1954,BedH1955,By1957,Tr1961}
for arbitrary excited states, which 
is the subject of the current paper.

According to Sec.~3 of 
Ref.~\cite{KlBaMaRi2002}, a suitable approximation
for the incorporation of strong-interaction
effects into the discussion of energy
levels of electromagnetically bound particle-antiparticle systems
consists in the addition of an 
intense, short-range (radius $a_h$) potential,
which can be assumed to approximate a 
hard-sphere potential for $r \leq a_h$.
Such intense, short-range hard-sphere
(infinitely intense) potentials, when added to a long-range
Coulomb potential, cannot be treated on the 
basis of Wigner--Brillouin perturbation theory
(for a recent overview, see Chap.~5 of Ref.~\cite{JeAd2022book}).
This is because the expectation value of 
a hard-sphere potential (which assumes an infinite
value inside the core) is infinite for
any bound state in a Coulomb potential. Rather,
one needs to apply so-called radius perturbation
theory where the expansion parameter is
the ratio $a_h/a_0$ of the radius $a_h$ of the hard-sphere
potential and the Bohr radius~$a_0$.

The calculation of the shift $E_n \to E_n + \delta E$
due to the hard-sphere potential has been considered
by Deser {\em et al.} (see Ref.~\cite{DeGoBaTh1954}) and
Trueman (see Ref.~\cite{Tr1961}),
and the result is therefore commonly referred to as the 
Deser--Trueman formula (see 
Refs.~\cite{BaDe1999,MiIv2001,De2001scat,YaEtAl2009,%
NuSuYaFa2009,BoHivD2010,RiFa2017,MoHoRi2020}.
It is generally assumed 
[see, {\em e.g.}, Eq.~(3.69) of Ref.~\cite{KlBaMaRi2002}]
that the energy perturbation takes the form 
\begin{equation}
\label{basic_assumption}
\delta E = f(n, \ell) \, \left( \frac{a_h}{a_0} \right)^{2 \ell + 1} \,,
\end{equation}
where $f(n, \ell)$ remains to be determined.

The problem of the calculation of $\delta E$
arises in hadronic bound systems such as
protonium~\cite{Ba1989,CaRiWy1992,BaFrGa2001,KlBaMaRi2002}, 
and in other bound systems where hadrons are the 
constituent particles.
(The concept of the Bohr radius finds 
a natural generalization in these systems
as $a_0 = \hbar/(\alpha m_r c)$, 
where $m_r$ is the reduced mass of the 
two-body bound quantum system.)
It is well known that the nucleon-antinucleon 
potential is more attractive than the nucleon-nucleon
potential, in view of the fact that $\omega$-meson 
exchange flips sign and becomes attractive~\cite{Ri2020}.
For Coulombic bound systems where the constituent
particles are hadrons  and annihilation is involved, this means that, 
for very close approach $r \leq a_h \sim 1\, {\rm fm}$, 
the strong interaction leads 
to immediate annihilation of the 
particles. The wave function of the Coulombic
bound system thus needs to vanish inside 
the range of the strong interaction, typically
assumed to be of the order of $a_h \sim 1\, {\rm fm}$.
Therefore, the strong-interaction potential,
on the scale of the Bohr radius, acts,
somewhat counter-intuitively, as a hard-sphere
potential. One thus needs to calculate the 
perturbed energies under the assumption that the 
wave functions vanish for $r < a_h$, as induced 
by the potential $V_h$.
Hence, again, somewhat counter-intuitively, the 
strong-interaction correction shifts the 
bound-state energies of protonium upwards in energy,
i.e., it reduces the modulus of the binding energy.

The necessity to calculate $\delta E$ 
has been recognized in the early days of the 
analysis of mesic atoms~\cite{DeGoBaTh1954,By1957,Tr1961,Ba1989,KlBaMaRi2002}.
Our goal here is to derive a formula for 
$\delta E$ which is valid for general
quantum numbers $n$ and $\ell$,
and to verify the result based on 
numerical calculations (see Sec.~\ref{sec2}).
The latter require the use of extended-precision
arithmetic, because, for higher $\ell$, 
the perturbations $\delta E$ becomes
very small in view of the their
scaling with $(a_h/a_0)^{2 \ell + 1}$
(see Sec.~\ref{sec3}).
Conclusions are drawn in Sec.~\ref{sec4}.

\section{Derivation}

\subsection{Mathematical Foundations}

Let us now discuss the mathematical details;
these are straightforward but worthy to be recalled, for reference.
We consider the physical system where
a short-range hard-sphere potential $V_h(r)$
perturbs an attractive  Coulomb potential $V_C(r)$,
\begin{equation}
V_h(r) = \left\{ 
\begin{array}{cc}
\infty  & \qquad r \leq a_h  \\
0 & \qquad r >  a_h 
\end{array}
\right. \,,
\qquad
V_C(r) = -\frac{e^2}{4 \pi \epsilon_0 r} = 
- \frac{E_h a_0}{r} \,.
\end{equation}
Here, $e$ is the electron charge, $\epsilon_0$ is
the vacuum permittivity, 
$c$ is the speed of light, and
$\hbar$ is Planck's unit of action.
One notes that, by constrast, the perturbative
potentials induced by the nuclear-finite-size 
effect remain finite near the origin and allow 
for a perturbative treatment (see Chap.~5 of Ref.~\cite{JeAd2022book}).
The infinite character of the potential $V_h$ makes the 
current treatment more challenging.
The (generalized)
Hartree energy is $E_h \equiv m_r c^2 \alpha^2$, 
and $a_0$ is the Bohr radius
as defined earlier. It is interesting to note that 
discontinuities in binding potentials,
such as the one encountered for $V_h(r)$ at
$r = a_h$, are not unique to 
strong-interaction corrections, but 
also occur in the context of 
spherical quantum reflection traps~\cite{JuRo2008}.
For an exotic two-body bound state, 
the Bohr energy levels $E_n$ are given by
\begin{equation}
\label{En}
E_n = -\frac{E_h}{2 n^2} = -\frac{\alpha^2 m_r c^2}{2 n^2} \,.
\end{equation}
The bound-state 
wave functions are well known to have the 
form $\psi_{n \ell m}(\vec r\,) = R_{n \ell}(r) \, Y_{\ell m}(\theta, \varphi)$,
where $Y_{\ell m}(\theta, \varphi)$ is the spherical harmonic 
and the radial wave function $R_{n \ell}(r)$ 
depends on the principal quantum number $n$ 
and the orbital angular momentum $\ell$ as follows
(we recall the formula for convenience)
\begin{equation}
\label{radial}
R_{n\ell}(r) =
\frac{\sqrt{(n - \ell - 1)!}}{\sqrt{(n + \ell)!}} \,
\frac{2^{\ell + 1}}{(a_0)^{3/2} \, n^2} \,
\left( \frac{r}{a_0 \, n} \right)^\ell \,
\exp\left( -\frac{r}{a_0 n} \right) \;
L^{2\ell+1}_{n-\ell-1} \left(\frac{2 r}{a_0 n} \right) \,.
\end{equation}
Here, $L_k^s(x)$ is the associated Laguerre polynomial.
Problems arise because the expectation value
$\langle V_h \rangle = 
\langle n \ell m | V_h | n \ell m \rangle$
diverges for any quantum numbers $n$, $\ell$ and $m$,
because even for higher $\ell$, the 
wave functions are not completely vanishing 
inside the hard-sphere radius $a_h$, 
no matter how small $a_h$ is.
Hence, ordinary perturbation theory is 
not applicable. 
A suitable expansion parameter is the ratio 
of the radius $a_h$ of the hard core
to the Bohr radius $a_0$
(see Refs.~\cite{DeGoBaTh1954,BedH1955,By1957,Tr1961,%
Ba1989,CaRiWy1992,BaFrGa2001,KlBaMaRi2002}).

\begin{figure}[t!]
\begin{center}
\begin{minipage}{0.99\linewidth}
\begin{center}
\includegraphics[width=0.91\linewidth]{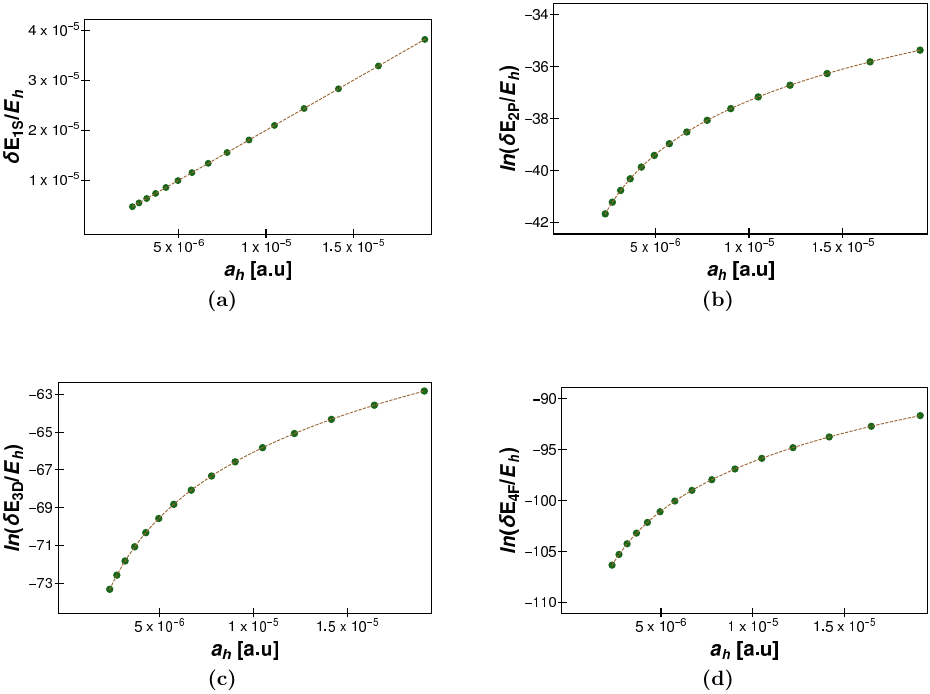}
\caption{\label{fig1} Numerical values for the 
energy shift $\delta E$ induced by a hard-sphere 
potential of radius $a_h$ are plotted, as 
a function of $a_h$, for a reference 
hydrogenic $1S$ state [panel (a)], $2P$ state [panel (b)], 
$3D$ state [panel (c)], and $4F$ state [panel (d)].
The dots are from the numerical approach outlined in 
Sec.~\ref{sec2}, while the dashed curve represents 
the analytic result given in Eq.~\eqref{res}.
Logarithmic scales are chosen for the ordinate
axes in panels (b), (c) and (d).
By ``a.u.'', we refer to atomic units,
{\em i.e.}, the radius $a_h$ is given in units of $a_0$.}
\end{center}
\end{minipage}
\end{center}
\end{figure}

%
%
\subsection{Radius Perturbation Theory}
\label{sec2}

After recognizing that any nonrelativistic hydrogenic 
bound-state wave function necessarily has the 
form $\psi(\vec r\,) = R_{n \ell}(r) \, Y_{\ell m}(\theta, \varphi)$,
the problem reduces to the calculation of a
bound-state solution that fulfills the 
matching condition $R_{n \ell}(r = a_h) = 0$
and approximates the solution~\eqref{radial}
of the radial wave equation.
We start from the representation of the 
radial function $R_{n \ell}(r)$ 
given in Eq.~\eqref{radial} in terms of a 
Whittaker $W$ function~\cite{WhWa1944,Ba1953vol1},
\begin{equation}
\label{Wh}
R_{n\ell}(r) =
\frac{ (-1)^{n + \ell + 1} \; W\left(n, \ell + \tfrac12, \frac{2 r}{n a_0} \right)}%
{ \sqrt{(n-\ell-1)!} \sqrt{ (n + \ell)! } \sqrt{a_0} n r } \,.
\end{equation}
The task is to adjust the principal quantum 
number $n$ from its integer value, which is attained 
for the unperturbed Schr\"{o}dinger--Coulomb problem,
to a non-integer $\nu$, chosen so that 
after finding $\nu$, close to $n$, the wave function vanishes
when $r = a_h$
\begin{equation}
W\left(\nu, \ell + \tfrac12, \frac{2}{\nu} \frac{ a_h}{ a_0} \right) = 0 \,,
\end{equation}
and the energy shift is given by
\begin{equation}
\delta E = \left[ \left(\frac{n}{\nu}\right)^2 - 1 \right] \, E_n \,,
\end{equation}
where $E_n$ is the unperturbed energy given in Eq.~\eqref{En}.
By carefully considering the shift in the 
short-range $r \to 0$ asymptotics of the Whittaker
function~\cite{WhWa1944,Ba1953vol1,La1969} 
in Eq.~\eqref{Wh} upon a departure of $n$ from its
initial integer value, one finds the compact result
\begin{equation}
\label{res}
\delta E = \frac{2 \alpha_{n, \ell} \, \beta_\ell}{n^3} \,
\left( \frac{a_h}{a_0} \right)^{2 \ell + 1} \, E_h \,,
\qquad
\alpha_{n,\ell} = \prod_{s = 1}^\ell 
\left( \frac{1}{s^2} - \frac{1}{n^2} \right) \,,
\quad
\beta_\ell = \frac{2 \ell + 1}{[(2 \ell + 1)!!]^2} \,,
\end{equation}
where $E_h$ is the Hartree energy,
and $\alpha_{n,0} = 1$, consistent with the identification
of an empty product being equal to one. The square of the 
double factorial, $[(2 \ell + 1)!!]^2$,
in the denominator of Eq.~\eqref{res} leads to a 
fortunate suppression of the strong-interaction
shift for Rydberg states with high angular momenta.
Corrections to Eq.~\eqref{res} are one power 
of the ratio $a_h/a_0$ higher 
than the leading order~\cite{Tr1961}. 
In view of the fact that the generalized 
Bohr radius $a_0$ is proportional to $1/\alpha$,
the corrections can be expressed,
as explained in Ref.~\cite{GaLyRu2008},
in powers of $\alpha$.

Our result can easily be expressed in 
terms of a quantum defect $\delta$,
where the bound-state energy $E_n$ shifts to a 
value $E'_n$, with a shifted principal quantum number
$n-\delta$,
\begin{equation}
\label{En_defect}
E_n = -\frac{\alpha^2 m_r c^2}{2 n^2} \to
E'_n =  -\frac{\alpha^2 m_r c^2}{2 (n - \delta)^2} \,,
\qquad
\delta = - 2 \alpha_{n, \ell} \, \beta_\ell \,
\left( \frac{a_h}{a_0} \right)^{2 \ell + 1} \,.
\end{equation}
One notes that quantum defect theory is often used
in order to describe energy shifts due to ``core'' 
potentials~\cite{GrFaSt1979}.

%
%
\subsection{Comparison to the Literature}

For $S$ states with orbital angular momentum
$\ell = 0$, the result~\eqref{res} specializes
to the formula
\begin{equation}
\delta E_{nS} = -\frac{4}{n} \frac{a_h}{a_0}\, E_n \,,
\end{equation}
where $E_n$ is the Schr\"{o}dinger--Coulomb energy
given in Eq.~\eqref{En}.
This is in agreement with Eq.~(1.1) of Ref.~\cite{Tr1961}
if one identifies the hadronic size $a_h$ with 
the $S$-wave scattering length $a_{\ell = 0} \equiv A_0$ 
used in Ref.~\cite{Tr1961}.
For $P$ states ($\ell = 1$), 
the result~\eqref{res} implies that
\begin{equation}
\label{res_nP}
\delta E_{nP} = -\frac{4}{3 \, n} \, 
\left(1 - \frac{1}{n^2}\right) \, 
\left(\frac{a_h}{a_0}\right)^3 \, E_n \,,
\end{equation}
which is also in agreement with Ref.~\cite{Tr1961},
but it takes a bit more work to verify this conclusion.
Namely, one needs to realize that the quantity 
$\varepsilon_\ell$ defined in Eq.~(2.11) of Ref.~\cite{Tr1961},
which parameterizes the phase shift according to
Eq.~(2.13) of Ref.~\cite{Tr1961}, actually diverges
for a hard-sphere potential. This is easily seen 
if one takes into account that the complete radial wave function
$\Psi$ (in the notation of Ref.~\cite{Tr1961}) vanishes 
at the boundary of the hard-sphere potential, and hence,
the logarithmic derivative of the radial wave function,
$(1/\Psi) (\dd \Psi/\dd r)$, diverges at the boundary 
of the hard-sphere potential.
One then combines Eqs.~(4.4) and (5.7) of Ref.~\cite{Tr1961},
realizes that $R$ in the notation of Ref.~\cite{Tr1961}
is our $a_h$, while $B$ in the notation of Ref.~\cite{Tr1961}
is our $a_0$, and shows the agreement of Trueman's 
paper~\cite{Tr1961} with our analysis for $P$ states.

For $D$ states ($\ell = 2$), which were not thoroughly considered in 
Refs.~\cite{Tr1961,CaRiWy1992,BaFrGa2001,KlBaMaRi2002},
we obtain
\begin{equation}
\label{res_nD}
\delta E_{nD} = -\frac{4}{45 \, n} \,
\left(\frac14 - \frac{1}{n^2}\right) \, 
\left(1 - \frac{1}{n^2}\right) \, 
\left(\frac{a_h}{a_0}\right)^5 \, E_n \,.
\end{equation}
While the scaling with $(a_h/a_0)^5$ is in agreement
with the basic assumption given in Eq.~\eqref{basic_assumption},
the prefactor is nontrivial and drastically 
reduces the magnitude of the effect for 
Rydberg $D$ states. 
For example, the value of the 3D protonium energy shift
(expected to be roughly similar to the imaginary part of the shift), from
Table 3.4 of Ref.~\cite{KlBaMaRi2002}, was estimated as 
$\sim 2 \times 10^{-6} \, {\rm eV}$,
while our formula with the prefactor gives the significantly smaller
shift $\sim 8 \times 10^{-9} \, {\rm eV}$.

%
%
\section{Numerical Verification}
\label{sec3}

The use of exponential numerical grids 
for the calculation of Schr\"{o}dinger--Coulomb 
energy eigenvalues has been pioneered in Ref.~\cite{SaOe1989}.
The use of extended-precision arithmetic~\cite{julialang,Ba2015}
facilitates the calculation of precise eigenvalues
even under extreme conditions where the perturbations
of a binding potential (here, a Coulomb potential)
are minute. In the current section, we thus 
ask the academic question whether or not lattice methods can
be optimized to the point where 
our analytic result can be verified against
numerical calculations even in parameter ranges
where the numerical shifts are not detectable
by standard-precision arithmetic. For our calculations, 
in terms of an academic check, we 
thus use a lattice with $N = 20\,000$ points, 
a maximum radial distance $R_N = 280 \, a_0$, 
and a minimum distance $R_1 = R_N / \exp[ (N-1) \, \delta x] =
2.459 \times 10^{-24} \, a_0$,
where the lattice spacing parameter is $\delta x = 0.003$.
Lattice points are distributed according to 
the formula 
\begin{equation}
R_i = R_1 \, \ee^{ (i-1) \delta x} \, a_0  \,,
\qquad
1 \leq i \leq N \,.
\end{equation}
Numerically, we approximate the potential 
$V_h(r)$ as vanishing for $r > a_h$, while inside
the hard sphere, we assign a numerical value of 
\begin{equation}
V_\infty = 10^{32} \, E_h
\end{equation}
as an approximation to an infinite value.
We use extended-precision arithmetic of 1024 decimal 
digits, in order to overcome numerical instability
problems, and determine the energy eigenvalues 
by inverse iteration~\cite{Po1921,TrBa1997}.
In order to verify Eq.~\eqref{res},
we vary $a_h$ with a reference lattice 
point $a_h = R_s$, where $s$ is a lattice
index chosen so that $R_s$ is in the 
range of $10^{-6}$ to $10^{-5}$.
This parameter range serves as a testing
ground for numerical lattice methods.
For reference $F$ states, because the 
energy displacement is proportional to $(a_h/a_0)^7$,
this means that $\delta E$ needs to be determined 
with an accuracy of up to 47~decimals
in order to discern the effect of the hard-sphere 
correction to the Coulomb potential.
We rewrite the second derivative 
with respect to the radial coordinate,
which is inherent to the radial Schr\"{o}dinger--Coulomb
eigenvalue equation, in terms of a 
derivative with respect to the lattice index $i$,
and use an 11-point finite-difference approxmation
in the numerical calculations (see Chap.~20 of Ref.~\cite{KoSiBa2020}).

We have carried out extensive numerical checks of
the result~\eqref{res} for all hydrogenic states
with principal quantum numbers $1 \leq n \leq 4$.
Here, we concentrate on states with maximum 
angular momentum quantum $\ell$ for given 
principal quantum number $n$ (see Fig.~\ref{fig1}).
Let us add a numerical example.
For $a_h = R_{i=14\,500} = 1.9117 \times 10^{-5} a_0$
(we pick this value in order to illustrate the 
high precision achieved using our numerics),
and a reference $4F$ state, we obtain 
\begin{equation}
\left. \delta E(4F) \right|_{a_h = 1.91 \times 10^{-5} a_0}  
= 1.58 \times 10^{-40} \, E_h \,,
\end{equation}
in excellent agreement with Eq.~\eqref{res},
which implies that
\begin{equation}
\left. \delta E(4F) \right|_{a_h = 1.91 \times 10^{-5} a_0}
= \frac{1}{5\,898\,240}
\left( 1.91 \times 10^{-5} \right)^7 \, E_h
= 1.58 \times 10^{-40} \, E_h \,.
\end{equation}
One notes that the energy shift could not be numerically
determined with standard arithmetic which is typically limited 
to 32 \!decimal digits in Fortran quadruple
precision~\cite{XLFortranLinux}.

%
%
\section{Conclusions}
\label{sec4}

The problem treated in the current article 
possesses a compact formulation but, nevertheless,
has quite far-reaching consequences. Let us emphasize that the analytic and 
numerical calculations were performed for a hydrogenic 
reference system (in the non-recoil limit 
of an infinite mass of the nucleus), 
for definiteness and ease of notation.
In order to generalize them for 
general two-body bound systems (examples
include protonium), one sets $a_0$ equal to the generalized
Bohr radius $a_0 = \hbar/(\alpha m_r c)$,
where $m_r$ is the reduced mass.
The formula~\eqref{res} remains valid 
provided one reinterprets $E_h$ in terms
of the generalized Hartree energy 
$E_h = \alpha^2 m_r c^2$.
For systems with constituent particles 
of charge numbers $Z_1$ and $Z_2$,
one generalizes the result with the 
replacement $\alpha \to Z_1 \, Z_2 \, \alpha$.

We note that the 
Deser-Trueman formula reproduces measured strong interaction energy shifts
reasonably well.  For protonium 1S states, the formula gives a shift of 0.867 keV,
compared to the measured value 0.72(4) keV from Table 6.3 of Ref.~\cite{KlBaMaRi2002}.
And for systems not composed of particles and antiparticles, 
the Deser-Trueman with $a_h = a_{\ell = 0} = 1\,{\rm fm}$
(approximating the $S$-wave scattering length with a 
hard-sphere radius of $1\,{\rm fm}$)
still gives a rough estimate.  For example, for kaonic hydrogen ($K^- p^+$), the
formula suggests a strong interaction shift of 0.41 keV, while the measured shift
is 0.19(4) keV (from Table I of Ref.~\cite{YaEtAl2009}).
The validity of our formulas for the estimation of the 
strong-interaction correction for more general hadronic bound 
systems is supported by relatively recent investigations which
indicate that the scattering length, as compared to
protonium, is a decreasing function
of the atomic weight (see the conclusions of Ref.~\cite{PrBoLRZe2000}).

A few remarks might be in
order regarding the literature on the 
Deser--Trueman formula.
As already mentioned, the parameter $a_h$ 
finds a natural generalization as the $S$-wave
scattering length in the absence of the
Coulomb field.
With this idea in mind, it has been
stressed~\cite{Tr1961,KlBaMaRi2002} that the 
energy shift~\eqref{res}, for reference $P$
states, could be assumed to be proportional 
to a scattering ``volume'', namely,
to the third power of a parameter $R \approx a_0$,
which is commensurate with the scattering
length [see also Eq.~(5.7) of Ref.~\cite{Tr1961}].
For reference $D$ and $F$ states,
the energy shift is then proportional 
to the scattering ``hyper-volumes'',
{\em i.e.}, proportional to 
$(a_h)^5$ and $(a_h)^7$, respectively.
These assumptions are confirmed
in Eqs.~\eqref{res_nP} and~\eqref{res_nD}.
Our general result, given in Eq.~\eqref{res},
possesses a highly nontrivial prefactor,
which leads to a drastic suppression
of the effect for high angular momenta, beyond the 
simple scaling with $(a_h/a_0)^{2 \ell + 1}$.

Thus, our result allows for a reliable estimate of the 
strong-interaction correction to 
non-$S$ Rydberg levels of electromagnetically 
bound systems containing hadronic 
constituent particles, and is thus essential
for the reliability of theoretical
predictions for corresponding experiments 
involving exotic hadronic atoms.
Our investigations here focus on the estimation
of the numerically small strong-interaction correction for 
high-precision experiments in exotic, hadronic bound 
systems involing non-$S$ states (see also Appendix~\ref{appa}), 
where the presence of numerically small, but nonvanishing
strong-interaction shifts can be an obstacle to the 
detection of New Physics effects~\cite{BaEtAl2025,AdJe2025prr}.
Our general-purpose formula, given in Eq.~\eqref{res},
addresses the issue.

\section*{Acknowledgments}

\vspace{-0.2cm}

The authors acknowledge insightful discussions with Professor
Jean-Marc~Richard.  This work was supported by the National Science Foundation
through Grants PHY-2308792 (G.S.A.), PHY--2110294 and PHY--2513220 (U.D.J.),
and by the National Institute of Standards and Technology Grant 60NANB23D230
(G.S.A.).

\appendix

%
%
\section{Strong--Interaction Shifts: $\bm S$ versus non-$\bm S$ States}
\label{appa}

The focus in our current work is on Rydberg states
with manifestly nonvanishing angular momenta.
This approach constitutes a departure from
previous investigations on exotic bound systems,
which have primarily focused on measurements
of the ground-state parameters (energy and lifetime),
and its relation to $S$-wave scattering lengths
(see the comprehensive discussion 
in Refs.~\cite{GaLyRu2008,GaLyRu2009}).
Over the last three decades, there have been remarkable
efforts by the DIRAC collaboration at CERN in regard
to the $1S$ ground-state lifetime of pionium
($\pi^+\pi^-$ atom, see Refs.~\cite{DI2005,DI2015}),
and other bound states such as bound pion-kaon
systems~\cite{DI2017}. The pionic hydrogen collaboration
at the Paul--Scherrer--Institute (PSI) has focused
on pionic hydrogen measurements (see Ref.~\cite{pionicH,HeEtAl2014epja,HiEtAl2021})
and pionic deuterium (Ref.~\cite{HaEtAl1998}), in each
case, focusing primarily on ground-state properties.
The DEAR collaboration has determined 
ground-state properties of kaonic hydrogen~\cite{BeEtAl2005}, 
but has also, quite remarkably,
observed transitions from higher excited states of 
kaonic hydrogen to the ground state (the so-called $K_\alpha$,
$K_\beta$ and $K_\gamma$ lines, see Ref.~\cite{BeEtAl2005}).
Conversely, from the theoretical side, 
a lot of efforts have been undertaken to understand
the strong-interaction energy shift of the ground state of these 
exotic systems, and the strong-interaction contribution
to the ground-state lifetime~\cite{LyRu1996,GaGaLyRu1999,LyRu2000,KoRa2000,%
GaLyRuGa2001,LiLyRu2002,GaEtAl2002,MeRaRu2004,MeRaRu2005,MeRaRu2006,%
GaLyRu2008,GaLyRu2009,DoMe2011}).
The properties of the ground state of the 
exotic Coulomb systems offer tremendous insights 
into the strong-interaction contributions to the effective
Lagrangian describing the exotic bound systems. These
contributions can be absorbed, to good approximation, 
in contact terms, such as the 
last term on the right-hand side of Eq.~(4.2) of Ref.~\cite{GaLyRu2008}.
However, if one focuses on $S$ states of exotic systems,
then one overlooks the fact that the Rydberg states
(notably, those with nonvanishing angular momenta)
offer an almost ideal combination of very strong electromagnetic 
fields which approach the Schwinger limit (see also 
the pertinent discussion in Ref.~\cite{AdJe2025prr}),
while being largely free from strong-interaction effects.
They are thus very attractive candidates for a 
search for new physics~\cite{BaEtAl2025}.
For the latter states (Rydberg states with nonvanishing
angular momenta), the strong-interaction effects are
small and, in fact, vanishing within the approximations
considered in the treatment following Eq.~(5.27) of 
Ref.~\cite{GaLyRu2008}. However, if one would like to 
consider highly excited non--$S$
Rydberg states of exotic bound states,
then the approach based on contact terms 
involving coordinate-space Dirac--$\delta$ terms
in the effective Lagrangian is not applicable,
because the wave function of 
states with high angular momenta vanishes 
at the origin, while the strong-interaction 
correction does not vanish, while being parametrically
suppressed for the non--$S$ states. For this reason,
one needs an estimate of the strong-interaction shift
in order to discern possible tiny energy 
shifts due to residual strong-interaction shifts
from putative New Physics effects
in prospective measurements of Rydberg-state transitions
in heavy exotic systems~\cite{BaEtAl2025,AdJe2025prr}.
This is the issue addressed in the current investigation.

For $S$ states, within the approximation of treating
the strong-interaction correction 
as a hard-sphere potential (see Sec.~3 of
Ref.~\cite{KlBaMaRi2002}), our approach is 
consistent, but less precise, than the one
chosen in Refs.~\cite{LyRu1996,GaGaLyRu1999,KoRa2000,%
GaLyRuGa2001,LiLyRu2002,GaEtAl2002,MeRaRu2004,MeRaRu2005,MeRaRu2006,%
GaLyRu2008,GaLyRu2009,DoMe2011}.
Furthermore, it should be clearly stated
that our approach cannot, in principle, describe the 
spin-dependent terms of the nucleon-nucleon
interaction~\cite{DoRi1980,KoWe1986,MaTj1969,PaFrGi1982,%
WyGrNi1985,YaKoKoSu2008,LaCa2021,DuLaCa2023corr,DuLaDE2023}.
However, we reemphasize that it is our goal to 
present a universally applicable
approximation  of the strong-interaction
correction for non-$S$ Rydberg states,
where the strong-interaction correction is 
numerically so small that it 
is otherwise mostly ignored in the literature,
and utmost accuracy is not required for the 
treatment of the strong-interaction correction
[see the discussion following Eq.~(5.27) of
Ref.~\cite{GaLyRu2008}]. As a last remark,
the imaginary part of the $S$-wave scattering length 
is found to be commensurate with its real part
for all exotic systems considered in Refs.~\cite{LyRu1996,GaGaLyRu1999,KoRa2000,%
GaLyRuGa2001,LiLyRu2002,GaEtAl2002,MeRaRu2004,MeRaRu2005,MeRaRu2006,%
GaLyRu2008,GaLyRu2009,DoMe2011}).
While our hard-sphere approach only yields the 
real part of the energy correction for the non-$S$ Rydberg
states, which are at the focus of our interests, we can 
thus realiably estimate the imaginary part of the 
strong-interaction energy correction for the non-$S$ Rydberg
states to be commensurate with its real part.
In this context, it is useful to recall that the scattering lengths 
used for $S$-state calculations are 
manifestly complex rather than real quantities
(see Tables~7 and~8 of Ref.~\cite{GaLyRu2008}).
For $P$ states, the scattering volumes
obtained using optical nucleon-nucleon potentials
are, likewise, complex rather than real
(for example calculations, 
see Table~1 of Ref.~\cite{DuLaDE2023}).
For reference, we can point out that
the radiative photon-emission decay widths 
of excited atomic states (due to spontaneous emission
of photon and dipole-allowed transitions to 
energetically lower atomic bound states) are 
well known to be of order $\alpha^5 m_r$ (see Ref.~\cite{JeAd2022book}), 
uniformly for any orbital angular 
momentum. For $P$ states, the photon-emission decay width
is thus of the same order in $\alpha$ as compared to the 
hadronic decay widths described by the imaginary 
part of the strong-interaction correction for $P$
states (also of order $\alpha^5 m_r$). 
For excited $S$ states, the strong-interaction decay width
dominates (order $\alpha^3 m_r$), while for 
$D$ states and higher angular momenta, 
the strong-interaction decay width 
(order $\alpha^7 m_r$ and higher) is parametrically 
suppressed in comparison to the photon-emission decay width.

\end{document}